\documentclass[aps,prb,showpacs,twocolumn]{revtex4}
\usepackage{graphicx}

\begin{document}

%%%%%%%%%%%%%%%%%%%%%%%%%%%%%%%%%%%%%%%%%%%%%%%%%%%%%%%%%%%%%%%%%%%%%%%%%%%%%%%%%%%
\newcommand{\figureheight}{8.2 cm}
\newcommand{\putfig}[2]{\begin{figure}[h]
        \special{isoscale #1.bmp, \the\hsize \figureheight}
        \vspace{\figureheight}
        \caption{#2}
        \label{fig:#1}
        \end{figure}}

        % almost universal commands for equations and references
\newcommand{\eqn}[1]{(\ref{#1})}

\newcommand{\be}{\begin{equation}}
\newcommand{\ee}{\end{equation}}
\newcommand{\bea}{\begin{eqnarray}}
\newcommand{\eea}{\end{eqnarray}}
\newcommand{\bean}{\begin{eqnarray*}}
\newcommand{\eean}{\end{eqnarray*}}

\newcommand{\nn}{\nonumber}
%%%%%%%%%%%%%%%%%%%%%%%%%%%%%%%%%%%%%%%%%%%%%%%%%%%%%%%%%%%%%%%%%%%%%%%%%%%%%%%%%%%

%%% ----------------------------------------------------------------------

%%versione del
%%%%%%%%%%%%%%%%%%

%%%%%%%%%%%%%%%%%%

\title{Spin separation  in a T ballistic nanojunction  due to lateral-confinement-induced
spin-orbit-coupling}
\author{S. Bellucci$^a$, F. Carillo$^b$  and P. Onorato$^{a,c}$\\}
\address{
        $^a$INFN, Laboratori Nazionali di Frascati, P.O. Box 13, 00044 Frascati, Italy. \\
        $^b$NEST-INFM-CNR and Scuola Normale Superiore, I-56126 Pisa, Italy. \\
        $^c$Department of Physics "A. Volta", University of Pavia, Via Bassi 6, I-27100 Pavia, Italy.
}
\date{\today}

\begin{abstract}
We propose a new scheme of spin filtering employing ballistic
nanostructures in two dimensional electron gases (2DEGs). The
proposal is essentially based on the spin-orbit (SO) interaction
arising from the lateral confining electric field. This sets the
basic difference with other works employing ballistic crosses and T
junctions with the conventional SO term arising from 2DEG
confinement. We discuss the consequences of this different approach
on magnetotransport properties of the device, showing that the
filter can in principle be used not only to generate a spin
polarized current but also to perform an electric measurement of the
spin polarization of a charge current. We focus on single-channel
transport and investigate numerically the spin polarization of the
current.
\end{abstract}


\maketitle

%\section
{\em Introduction -} In recent years a great effort has been
devoted to the study and the realization of electric field
controlled spin based devices\cite{spintro}. Many basic building
blocks are today investigated theoretically and experimentally in
order to realize a fully spin based circuitry. Among them, a
particular relevance is covered by: (i) pure spin current
generation, (ii) voltage control of the spin polarization of a
current and (iii) the electric detection of this polarization. For
the same purpose many works have been focusing on the so called
spin Hall effect \cite{[3],[7],Hir,she2,she3,she4} and most of
the implementations in 2DEGs proposed for the spin manipulation
are mainly based on the spin-orbit (SO) interaction. The SO
Hamiltonian reads \cite{Thankappan}
\begin{equation}
\hat H_{SO} = -\frac{\lambda_0^2}{\hbar}e{\bf E}({\bf r})
\left[\hat{{\bf \sigma}}\times \hat{\bf p}\right]. \label{H_SO}
\end{equation}
Here ${\bf E}({\bf r})$ is the electric field,
 $\hat{{\sigma}}$ are
the Pauli matrices, { $\hat{\bf p}$ is the canonical momentum
operator  ${\bf r}$ is a 3D position vector }, $\lambda_0^2=
\hbar^2/(2m_0 c)^2$ and $m_0$  the  electron mass in vacuum. In
materials $m_0$ and $\lambda_0$ are substituted by their effective
values $m^*$ and $\lambda$. A significant SO term arises from the
interaction of the traveling charge carrier with strong electric
fields in solids. The SO term can be seen as the interaction of the
electron spin with the magnetic field, $B_{eff}$, appearing in the
rest frame of the electron.
%%%%%%%%%%%%%%%%%%%%%%%%%%%%%%%%%%%%%%%%%%%%%%%%%%%%%%%
%%%%%%%%%%%%%%%%%%%%%%%%%%%%%%%%%%%%%%%%%%%%%%%%%%%%%%%
\begin{figure}
\includegraphics*[width=.45\linewidth]{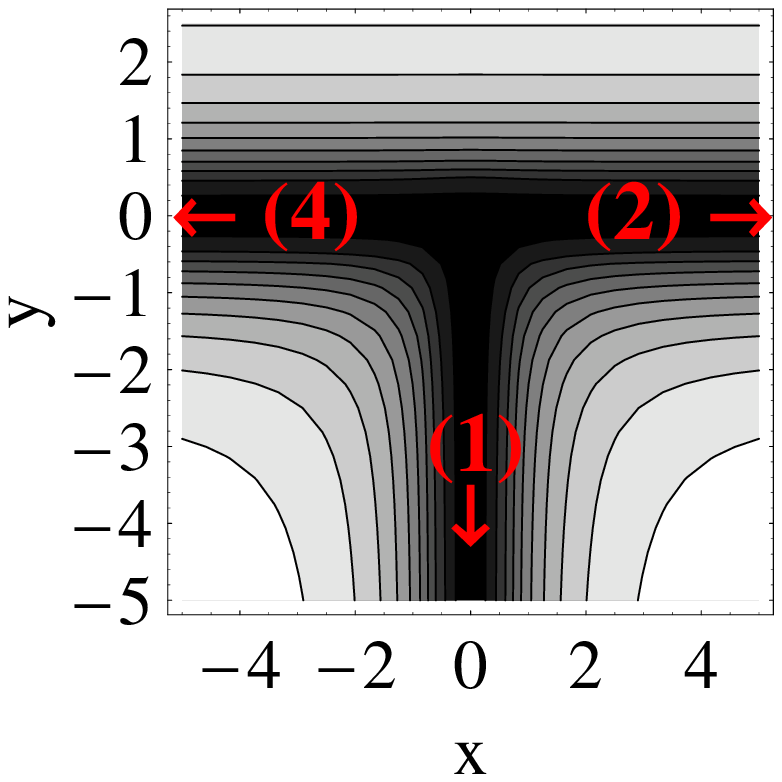}
\includegraphics*[width=.45\linewidth]{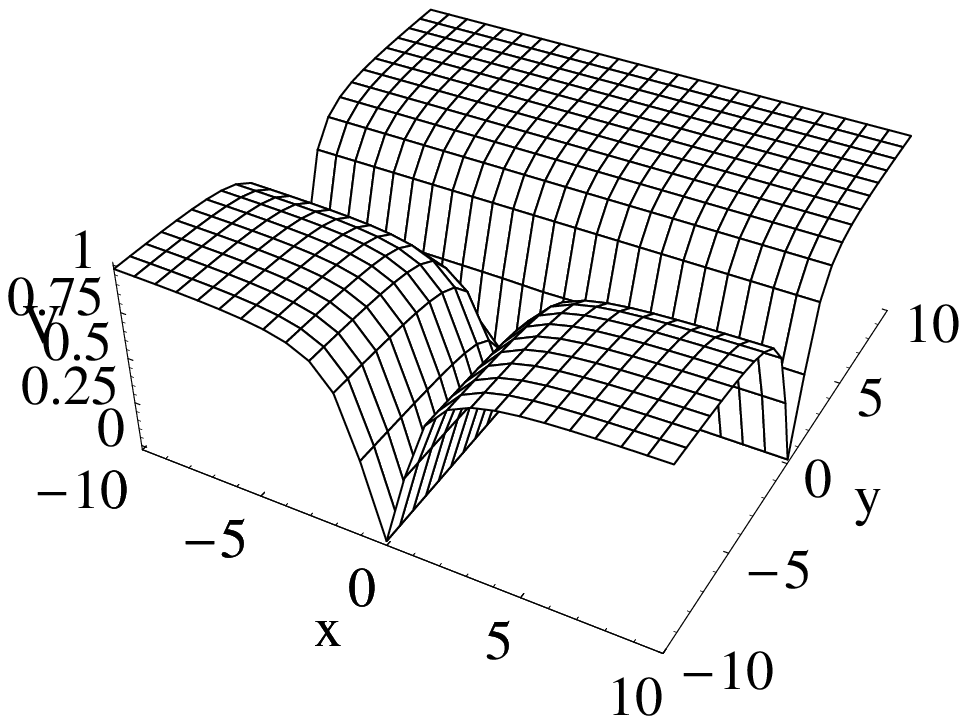}
 \caption {\label{fig1} : Density  and 3D Plots of the potential  $V_{c}(x,y)$ which
models a T-shaped
 conductor.
 This device   can be assumed
 as  a crossing junction between two quasi one dimensional wires of width $W$ which ranges from  $\sim
25$  nm  up to $100$ nm.}
\end{figure}
%%%%%%%%%%%%%%%%%%%%%%%%%%%%%%%%%%%%%%%%%%%%%%%%%%%%%%%
%%%%%%%%%%%%%%%%%%%%%%%%%%%%%%%%%%%%%%%%%%%%%%%%%%%%%%%

\

{\it Rashba Coupling -} { In the case of quantum heterostructures of
narrow gap semiconductors, a major contribution to the SO coupling
may originate intrinsically from its confining potential\cite{BR}.
The spin Hall effect in two-dimensional $2D$ electron systems
exploits the Rashba SO coupling ($\alpha$-coupling) due to an
asymmetry in quantum well potential that confines the electron
gas~\cite{Kelly}. The main component of the SO coupling  will be
along $\hat{z}$ and the Hamiltonian in Eq.~\ref{H_SO} will take the
form\cite{bonew} $ \hat H_{SO}^{\alpha} =
\frac{\alpha}{\hbar}\;\left[{\bf \sigma_x}{\bf p_y}-{\bf
\sigma_y}{\bf p_x} \right] $. $\alpha$ in vacuum is $\lambda_0^2 E_z
e$ while the highest value of $\alpha$ in 2DEGs is close to  to
$10^{-10}$ eV m as reported in refs.[\onlinecite{shapers,Schultz}].

The $\alpha$-SO coupling may generate a spin-dependent transverse
force on moving electrons\cite{1519,1519a,1519b,1519c}. This force
tends to separate different spins in the transverse direction as a
response to the longitudinal charge current, giving a qualitative
explanation for the Rashba spin Hall effect. In the presence of
Rashba SO coupling, however, the electron spin, particularly its
out-of-plane projection, is not conserved and, hence, the usual
continuity equation fails to describe the spin transport. This
makes the spin transport phenomena in this system rather
complicated.}

\

{\it Lateral-confinement-induced Coupling -} { Next we consider
low dimensional electron systems
formed by crossing Quantum Wires (QWs) %quasi-one-dimensional
%electron systems patterned in 2DEGs)
 through the analysis of the
SO coupling in a 2D electron system with an in-plane potential
gradient. In such systems a confining ($\beta$-coupling) SO term
arises from the in-plane electric potential that is applied to
squeeze the 2DEG into a quasi-one-dimensional
channel~\cite{Thornton,Kelly}. }

We adapt the general form of eq.(\ref{H_SO}) to the strictly 2D
case, where the degree of freedom of motion in the z direction is
frozen out ($\langle p_z\rangle=0$), and the potential energy,
$V_c$,  depends only on $x$ and $y$ coordinates. Then  the SO
Hamiltonian in this case can be written in the form\cite{morozb}:
\begin{equation}
\hat H_{SO}^\beta = \frac{\lambda^2}{\hbar} \hat{\sigma_z}
 \left[{{\nabla V_c(x,y)}}\times \hat{\bf
p}\right]_z. \label{Hb}
\end{equation}
The reduced Hamiltonian commutes with the spin operator
$S_z=\hbar/2\sigma_z$, and, hence, conserves spin. Thus a SO
coupling of this kind generates a spin-dependent force on moving
electrons while conserving their spins. The standard continuity
equation for spin density and spin current is naturally
established because of spin conservation.

 The
spin-conserving $\beta$-SO interactions are also at the basis of the
quantum spin Hall effects discovered
recently\cite{noish,qse,iii,hatt,noiq}.

 \

{\it Spin Filters - }In this paper we investigate  the spin
polarization of the current in the presence of  (spin conserving)
$\beta$-interaction in a T-shaped conductor, in particular we show
why a $\beta$-coupling scheme results in a different working
principle, as compared to equivalent structures exploiting
$\alpha$-coupling \cite{kk}.

In fact  {\it"spin filters"} based on the $\alpha$-coupling rely
on the precession of electrons spins during their motion through
the wave guides, while the scheme we propose here, based on the
$\beta$-coupling, preserves the $\hat{z}$ component of the spin at
the injection and sends electrons/holes to different stubs
according to their spin value. Differently from $\alpha$-coupling
schemes the device we propose should in principle allow one, for a
given charge current, to electrically measure the spin
polarization grade with its sign. As a first step to defend this
statement we write down the Hamiltonian of a T-stub with SO
$\beta$-coupling and make some qualitative considerations. In the
following step we extract a quantitative analysis of device's
transport properties using materials parameters from the
literature.

\

{\it$\beta$-SO coupling and effective magnetic field - } Here we
focus on the case of pure $\beta$-coupling.% that can be
%experimentally achieved by using a highly symmetric 2DEG, where
%$\alpha$ is tailored to zero\cite{koga}.

The basic building blocks of the nanojunctions that we discuss in the
following are the QWs.  The ballistic one-dimensional wire is a
nanometric  solid-state device in
 which the transverse motion (along $\xi$) is quantized into discrete modes,
 and the longitudinal   motion ($\eta$ direction) is free.
  In this case electrons are envisioned to propagate freely down a
  clean narrow pipe and electronic transport with no scattering can occur.

   In a Q1D wire, where  a parabolic lateral confining potential\cite{3840} along
$\xi$ ($\xi\equiv x$ for leads $1$  and $\xi\equiv y$ for leads $2$
and $4$) with force $\omega_d$ is considered ($V({\bf r}) \equiv
V(\xi) =\frac{ m^*\omega_d^2}{2}\xi^2$) it follows
\begin{equation}
\hat{H}_{SO}^\beta = \frac{\beta}{\hbar}\frac{\xi}{l_\omega}
\left(\hat{{\sigma}}\times \hat{\bf p}\right)_\xi \simeq
i\beta\frac{x}{l_\omega}\sigma_z \frac{\partial}{\partial \eta},
\label{H_b}
\end{equation}
where $l_\omega=(\hbar/m^*\omega_d)^{1/2}$ is the typical spatial
scale and  $\eta$ is the other direction in the 2DEG ($\eta \perp
\xi$). Thus, as we discussed in a previous paper\cite{noish},  in a
QW
 a uniform {\it effective magnetic  field}, $B_{eff},$ is present along
 $z$,
\begin{equation}
\widetilde{B}_{eff} = \frac{\lambda^2}{\hbar}\;{m^*}^2\omega_d^2
c\equiv\frac{\beta}{\hbar l_\omega}\frac{m^*c}{e}.\label{Beff}
\end{equation}
Then an electron of spin $S_z=s\hbar/2$ flowing in the QW
perceives in its rest frame a magnetic field $B_{eff}$, directed
upward or downward according to the sign of $s$. This results in
an interesting behavior of junctions between two wires, such as T
stubs and cross junctions, when a large enough $\beta$-coupling is
considered.

\

The discussion reported above for a QW  can be generalized to any
device patterned in a 2DEG.  The Hamiltonian of an electron moving
in a 2D device defined by a general confining potential $V_c(r)$
in which the $\alpha$-SO term is negligible can be written as \bea
H&=&\frac{\bf{p}^2}{2 m^* }+\frac{\lambda^2}{\hbar}e\left({\bf
E}(\bf r)\times {\bf p}\right)_z \sigma_z+V_c({\bf r})\nonumber \\
&=& \frac{{\bf \pi}^2}{2}+V_c({\bf r}) -\frac{\lambda^4 m^*}{2
\hbar^2}{e^2\left|{\bf E}(\bf r)\right|^2},\label{hhh} \eea where
$ \pi_i=(p_i-\epsilon_{ijz}\frac{\lambda^2}{\hbar}m^* e
E_j\sigma_z)$  and ${\bf E}({\bf r})={\bf \nabla} V_c({\bf r})$.

The commutation relation, \bea \nonumber \left[\pi_x,\pi_y\right]=-i
\hbar \left(\frac{\lambda^2}{\hbar}m^* e {\bf \nabla}\cdot{\bf
E}\right) \sigma_z \equiv  -i \hbar \frac{e}{c}{B}_{eff}({\bf
r})\sigma_z \eea is equivalent to that of a charged particle in a
transverse magnetic field, but here the sign of ${B}_{eff}({\bf r})$
depends on the direction of the spin along $\hat{z}$. It follows
that electrons with opposite spin states are deflected into opposite
terminals by a spin-dependent Lorentz force: \bea {\bf
F}=m^*\ddot{\bf r}(t)&=&-{\nabla V_{c}({\bf r})}+ \frac{e}{m^*
c}\left({\bf B}({\bf r})\times {\bf \pi} \right), \eea where ${\bf
B}({\bf r})\equiv s {B}_{eff}({\bf r})\hat{z}$ is a spin dependent
inhomogeneous magnetic field, with $s=\pm1$.

\

{\it  The T junction -}In a cross junction sample, the confining
electrostatic potential  for an electron is not exactly known.
However, it is plausible that there has to be a potential minimum
at the center of the junction. In this respect, it would be
appropriate to qualitatively model the smooth confining potential,
displayed in Fig.(1), of a T-stub structure as \bea V_T(x,y)&=&
\frac{m^*}{2}\omega_d^2 R^2 \frac{x^2
y^2}{(R^2+x^2)(R^2+y^2)}\vartheta(-y)\nn
\\  &+&
\frac{m^*}{2}\omega_d^2 R^2 \frac{y^2}{(R^2+y^2)}\vartheta(y), \eea
 where   $\vartheta(y)$ is a regular function which approximates
the step function ($\vartheta(y)\sim (1+\tanh(y/\rho))/2$ with
$\rho\ll l_\omega$). Here $R$ represents the effective radius of
the crossing zone, while $l_\omega$ can be related to the
effective width of the wires $W$, which is known to be smaller
than the lithographically defined one and can be further reduced
by using etched side gate electrodes.
  This technique also works on small gap
semiconductors such as InGaAs\cite{carillo} featuring small Schottky
barrier with metals.
 In general, one can relate the
frequency $\omega_d$ to $W$ as $\omega_d \sim
\frac{(2\pi)^2}{2}\frac{\hbar}{m^* W^2}$. This expression can be
obtained by comparing the energy levels of a harmonic oscillator
to those of a square potential well.

Far from the crossing zone, the confining potential describing the
wires reads  $V_c(x,-\infty)\sim \frac{m^*}{2}\omega_d^2 x^2$ or
$V_c(\pm\infty,y)\sim \frac{m^*}{2}\omega_d^2 y^2$. Thus,
asymptotically $B_{eff}$ is  given by Eq.~\ref{Beff}. To have an
idea of the strength of this magnetic field, we compare the
cyclotron frequency $\omega_c = e\widetilde{B}_{eff}/(m^*c)$ with
$\omega_d$,

\begin{equation}
\widetilde{B}_{eff}\sim \frac{(2\pi)^4}{4}\frac{\hbar c}{e}
\frac{\lambda^2}{W^4}\leftrightarrow
\frac{\omega_c}{\omega_d}\sim\frac{(2\pi)^2}{2}
\frac{\lambda^2}{W^2}.\label{Beff2}
\end{equation}

We report all our results as a function of the ratio
$\omega_c/\omega_d$. In numerical calculation $\omega_c/\omega_d$
takes values that are in the range defined by experiments on 2DEGs.
We estimate the effective value of $\lambda$ in the 2DEG from the
measured value of $\alpha$ in the literature\cite{shapers,HgTe} and from
the calculated band diagram of the same structures.
%This estimate does not consider the contribution to $\alpha$ deriving from the different value of the wave function at the two interfaces of the quantum well
In InGaAs/InP heterostructures $\lambda^2$ takes values
between $0.5$ and $1.5\,nm^2$. %(in agreement with the values used in ref.\cite{morozb} where   $W\sim 200 nm$).
For GaAs heterostructures $\lambda^2$ is one order of magnitude less
than in InGaAs/InP, whereas  for HgTe based heterostructures it can
be more than three times larger \cite{Schultz,HgTe}. Since the
lithographical width of a wire defined in a 2DEG can be as small as
20nm \cite{kunze}, we assume that $\omega_c/\omega_d$ runs from $1
\times 10^{-6}$ to $1\times10^{-1}$. In any case $W$ should be
larger than $\lambda_F$, so that at least one conduction mode is
occupied.

\

{\it Ballistic transport and calculations-} Here we report a
numerical study, limited to a single channel transport, i.e. we
assume that just the lowest subband of the QW is activated.
 When the characteristic sizes
of semiconductor devices are smaller than the elastic mean free
path of charge carriers, the carrier transport becomes ballistic.
It follows that the transport can be studied starting from the
probability of transmission from a probe to another one
following the B\"uttiker-Landauer formalism\cite{BL}.

The calculation of the transmissions amplitude  is based on the
simulation of classical trajectories of a large number of
electrons with different initial conditions. We want to determine
the probability $T^{s,s'}_{1j}$ of an electron with spin $s$ to be
transmitted to lead $j$ with spin $s'$ when it is injected in lead
$1$. This coefficient can be determined from classical dynamics
of electrons injected   at $y_0=- 7.5\; l_\omega$ (emitter
position) with an injection probability following a spatial
distribution $ p_0(x_0,y_0)\propto e^{-\frac{x_0^2}{l_\omega^2}}$
as in ref.[\onlinecite{gei92}]. The total energy $\varepsilon$ of
a single electron is composed by the free electron energy
$\varepsilon^0_y$ for motion along $y$ and the energy of the
transverse mode considered $\varepsilon^0_x$ due to the parabolic
confinement ($\varepsilon_x = \hbar\omega_d/2$ for the lowest
channel).

Thus, we have calculated $T^{s,s'}_{ij}$ determined by numerical
simulations of the classical trajectories injected into the
junction potential $V_c$  with boundary conditions\cite{noiq}
${\bf r}(0)\equiv (x_0,y_0)\;;{\bf v}(0)\equiv {\bf v}_0$, each
one with a weight $p_0(x_0)$. In general these transmission
amplitudes can depend on the position of the collectors along the
probes. In this paper we take into account $N_t=804$ classical
trajectories for each value of the parameters.

 \

Before the discussion about our results we want to point out that a
comparison involving theoretical and experimental results  allowed
us to test our approach. In fact in ref.[\onlinecite{noiq}] we
investigated the effects on the X-junction transport due to a quite
small external magnetic field, $B_{ext}$, by focusing on the so
called {\it quenched region}. The measured  "quenching of the Hall
effect"\cite{fordbvh} is  a suppression of the Hall resistance or "a
negative Hall resistance" $R_H$ for small values of  $B_{ext}$. The
results reported in ref.[\onlinecite{noiq}] showed a good agreement
with the experimental data thus confirming the reliability of our
approach.

%%%%%%%%%%%%%%%%%%%%%%%%%%%%%%%%%%%%%%%%%%%%%%%%%%%%%%%
%%%%%%%%%%%%%%%%%%%%%%%%%%%%%%%%%%%%%%%%%%%%%%%%%%%%%%%
\begin{figure}
\includegraphics*[width=.31\linewidth]{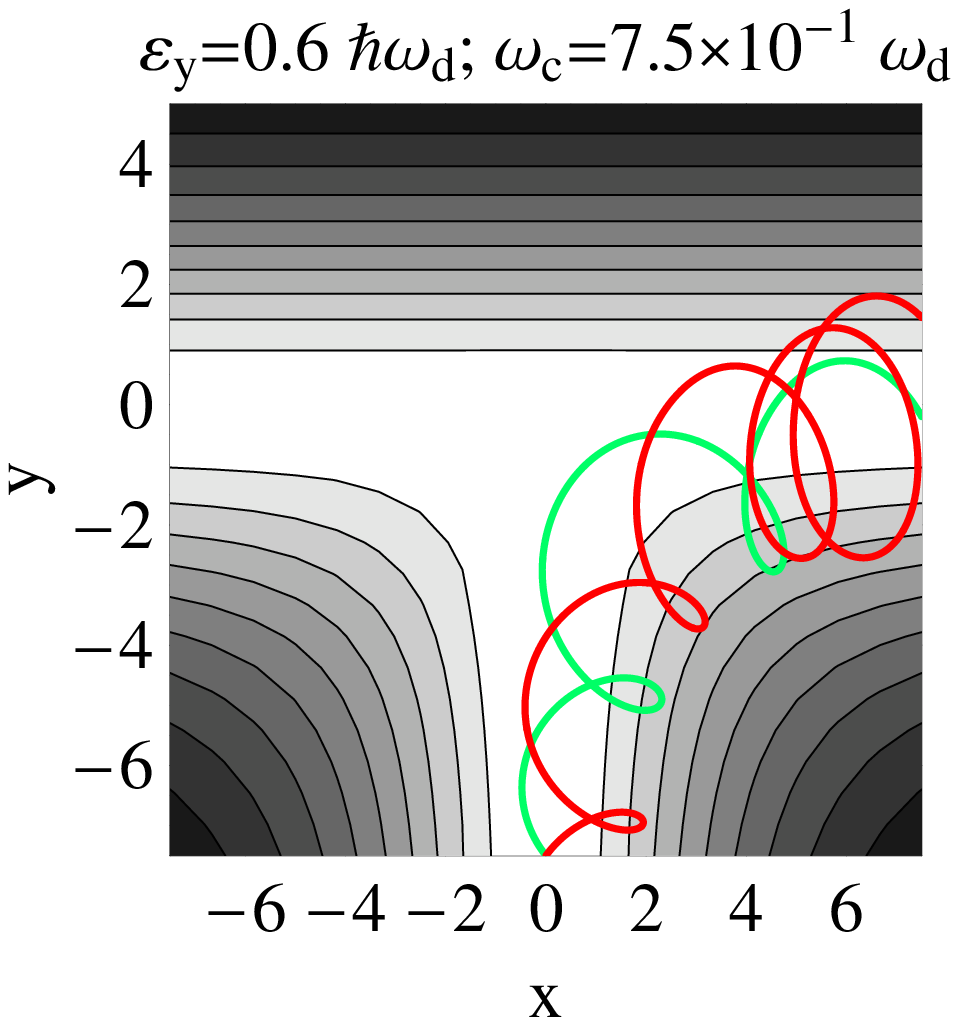}
\includegraphics*[width=.31\linewidth]{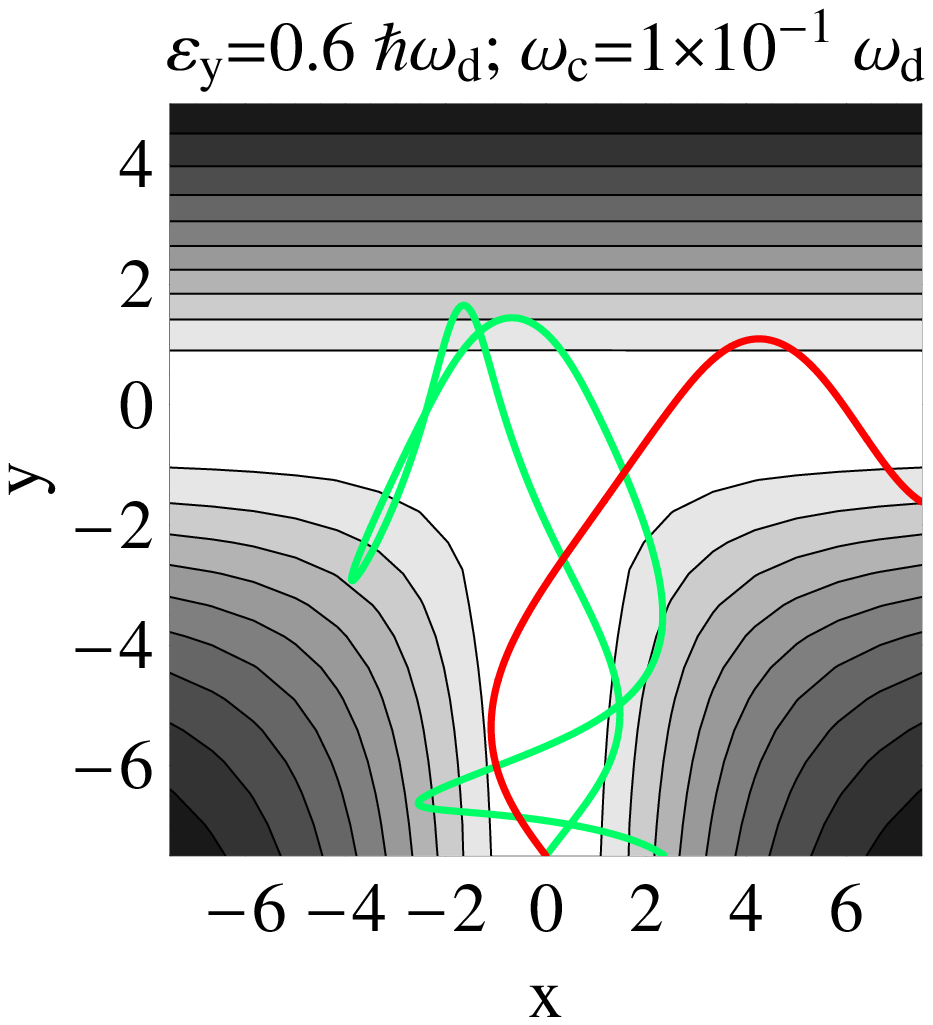}
\includegraphics*[width=.31\linewidth]{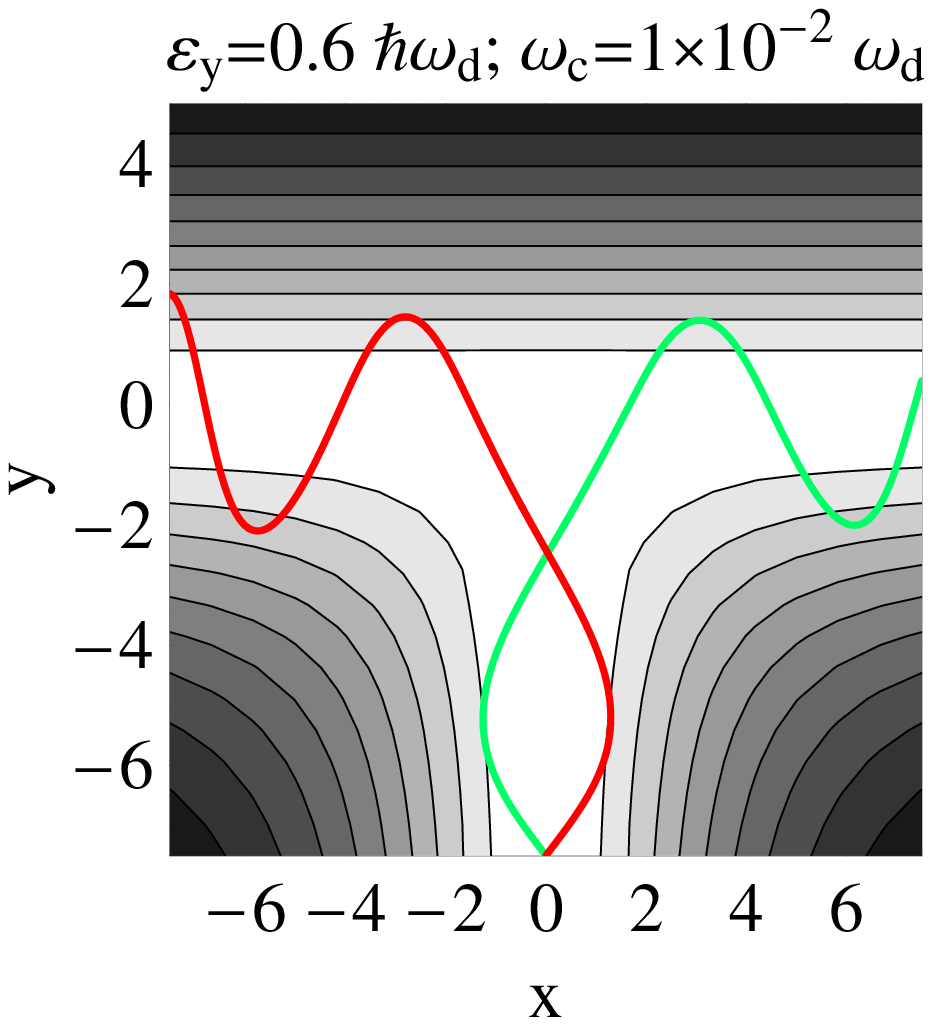}
 \caption {\label{fig3} Trajectories of the charges in the T-junction without SO.
 Each panel is for a different value of the external magnetic field. Red and green
 curves correspond to trajectories of the electrons injected in the lead 1 with the same
 energy and opposite $v_x$.}
\end{figure}
%%%%%%%%%%%%%%%%%%%%%%%%%%%%%%%%%%%%%%%%%%%%%%%%%%%%%%%
%%%%%%%%%%%%%%%%%%%%%%%%%%%%%%%%%%%%%%%%%%%%%%%%%%%%%%%
In order to show how the symmetry breaking produces a transverse
current $I_H$, we shortly discuss   the case of a T-shaped
junction, without SO term, in a uniform external magnetic field
$B$ directed along $\hat{z}$:
%shown in the bottom panel of Fig.~\ref{fig2}, is comparable to that
%reported in other theoretical\cite{noiq} and
%experimental\cite{fordbvh} works.
in Fig.~\ref{fig3} we report the corresponding classical electron
trajectories. The increase of the magnetic field results in a
broken symmetry between leads 2 and 4 and makes the probabilities
of transmission in the two leads very different.

The current $I_i$ at lead $i$ of a multi-probe device can be
expressed in terms of chemical potentials $\mu_j=eV_j$ at each lead
and of the transmission coefficient $T_{ij}$ as $I_i = e^2/h\sum_j
T_{ij}(V_i-V_j)$; normalization requires $\sum_j T_{ij}=1$
\cite{BL,gei92}. Thus to an injected current $I_0$ in lead $1$, it
corresponds a transverse Hall current
$$I_H=(T_{12}-T_{14})I_0.$$
This Hall current is mainly due to the electric field $\nabla
V_c({\bf r})$ for $y>0$ and to the broken symmetry between leads $4$
and $2$ due to the magnetic field.

%%%%%%%%%%%%%%%%%%%%%%%%%%%%%%%%%%%%%%%%%%%%%%%%%%%%%%%
%%%%%%%%%%%%%%%%%%%%%%%%%%%%%%%%%%%%%%%%%%%%%%%%%%%%%%%
%%%%%%%%%%%%%%%%%%%%%%%%%%%%%%%%%%%%%%%%%%%%%%%%%%%%%%%
\begin{figure}
\includegraphics*[width=.951\linewidth]{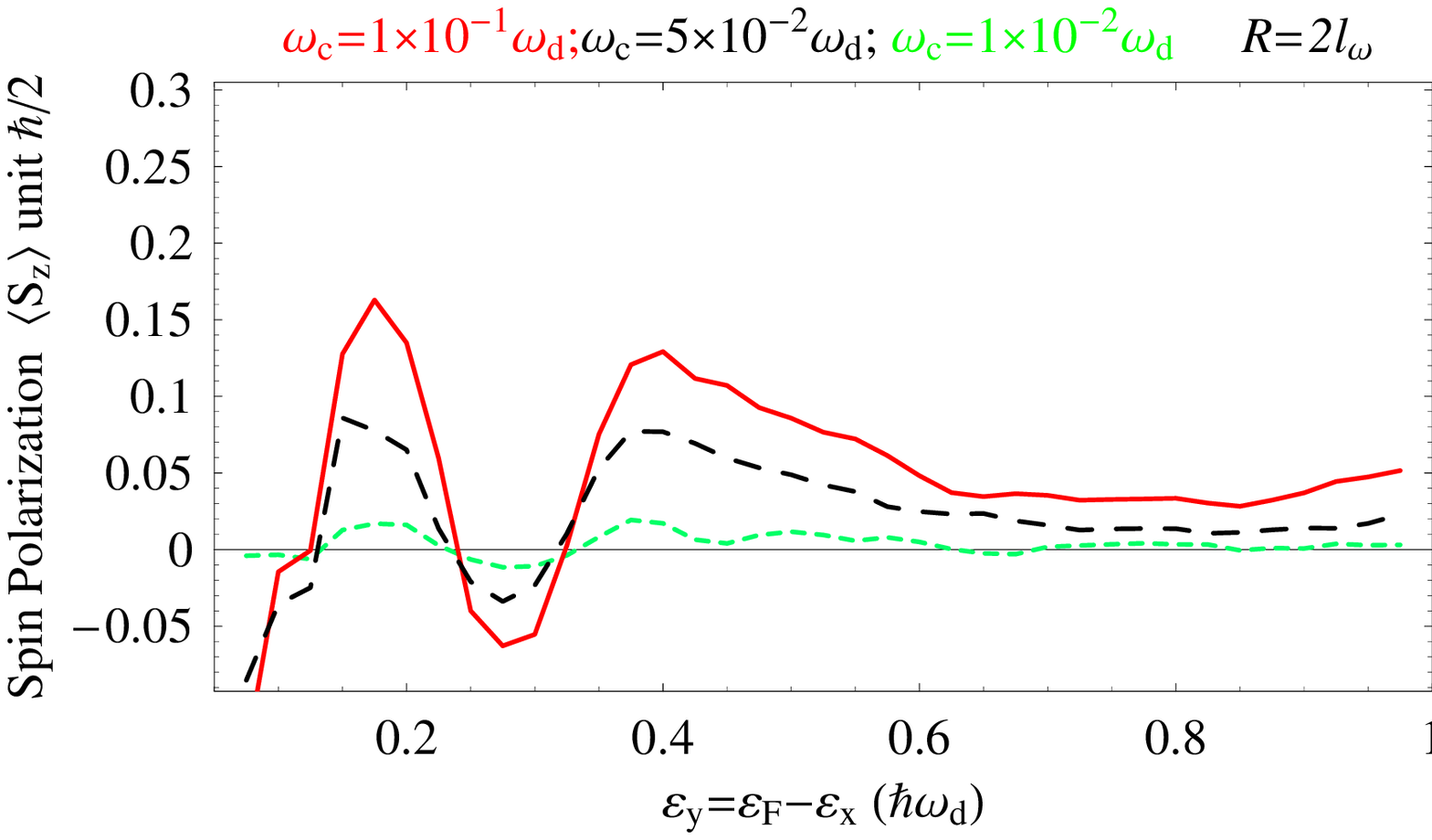}
\includegraphics*[width=.951\linewidth]{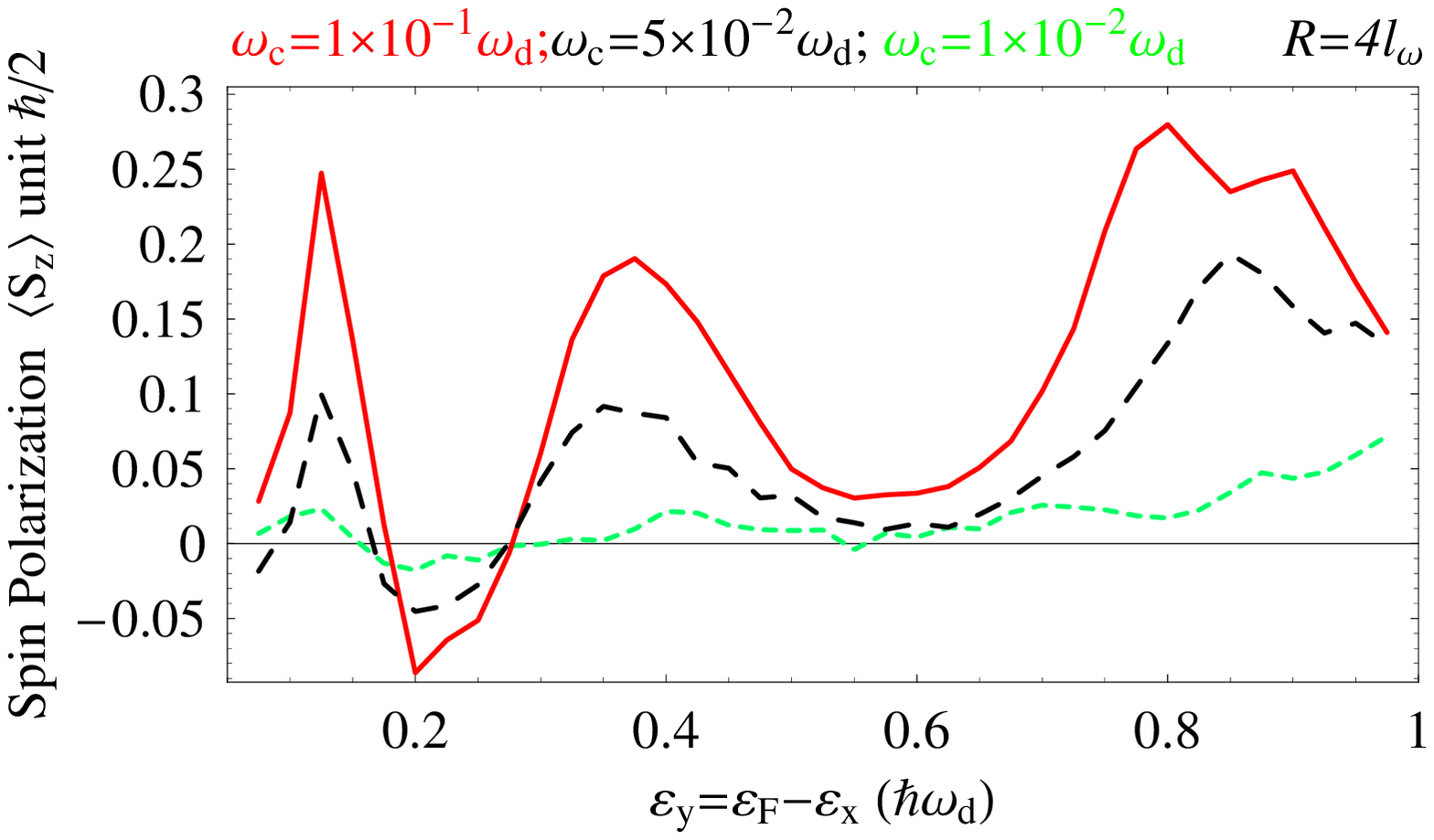}
\includegraphics*[width=.951\linewidth]{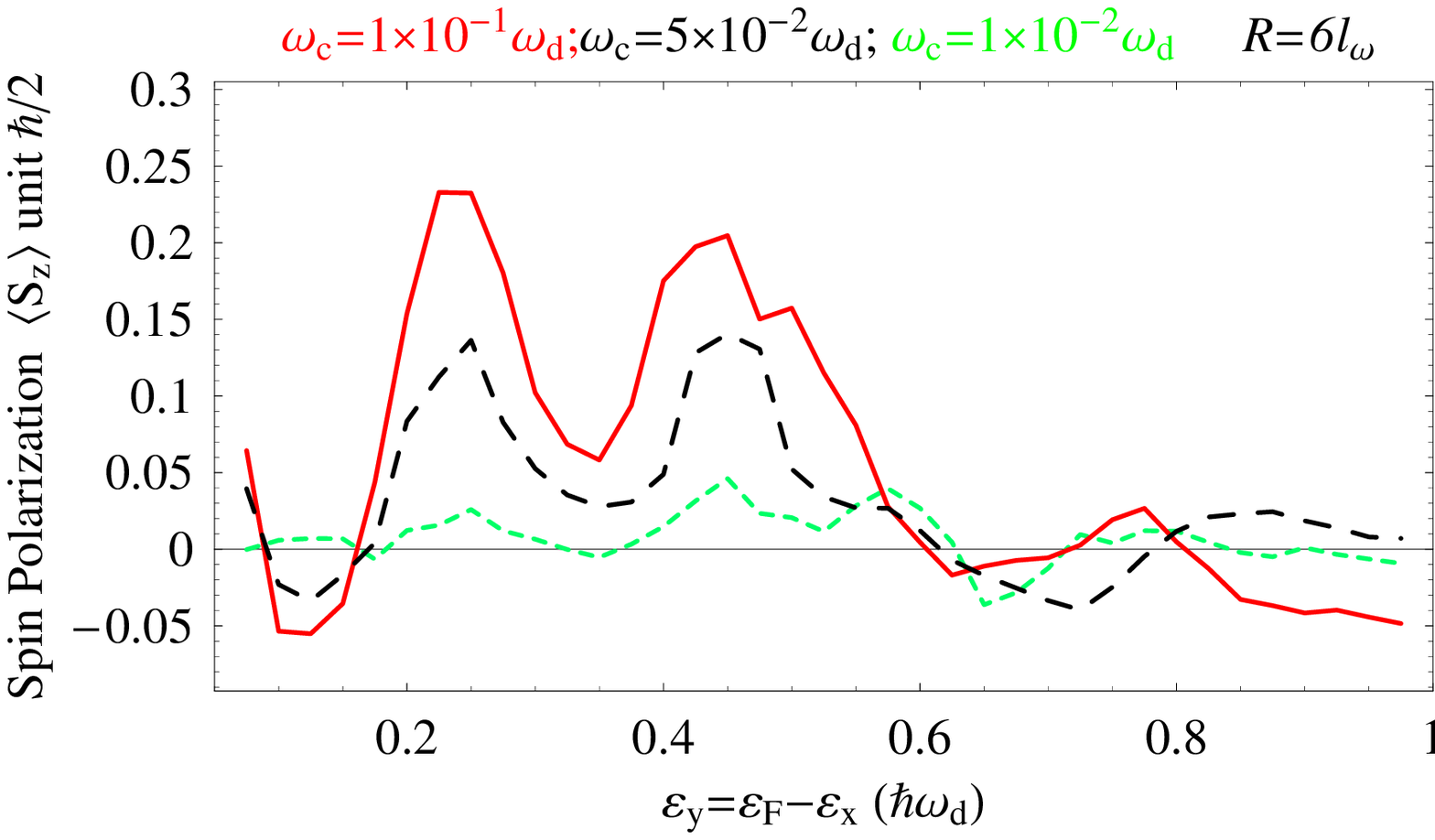}
 \caption {\label{fig4} Spin polarization at lead 2 $\langle S_z \rangle$ as a function of
the Fermi energy $\varepsilon_F$ in the case of zero external
magnetic field and $\beta$ SO coupling for three different values
of the radius $R$ of the crossing zone. Around $\varepsilon_y=0.2
\hbar \omega_d$ there is a quenching region corresponding to an
inversion of the spin polarization\cite{noiq}.}
\end{figure}
%%%%%%%%%%%%%%%%%%%%%%%%%%%%%%%%%%%%%%%%%%%%%%%%%%%%%%%
%%%%%%%%%%%%%%%%%%%%%%%%%%%%%%%%%%%%%%%%%%%%%%%%%%%%%%%

%%%%%%%%%%%%%%%%%%%%%%%%%%%%%%%%%%%%%%%%%%%%%%%%%%%%%%%
%%%%%%%%%%%%%%%%%%%%%%%%%%%%%%%%%%%%%%%%%%%%%%%%%%%%%%%
\begin{figure}
\includegraphics*[width=.851\linewidth]{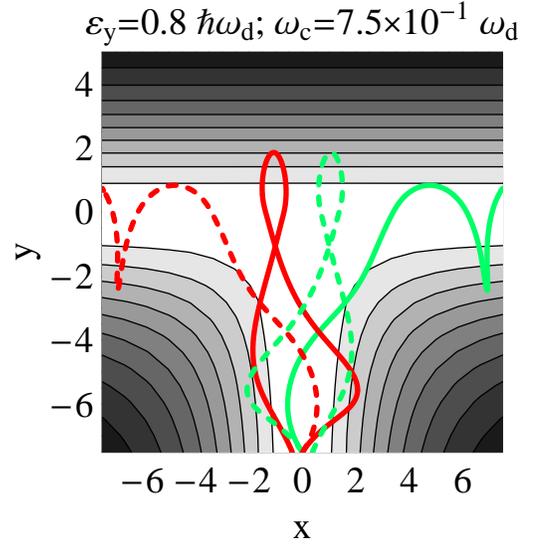}
 \caption {\label{fig5} Trajectory of an electron in a T junction
 with $\beta"$ SO term and vanishing external magnetic field.  Electrons are injected
 in lead 1 with a defined spin orientation along $\hat{z}$. Dashed
 lines correspond to $e^\downarrow$ solid lines to $e^\uparrow$.
 Electrons are injected with the same energy and opposite $\pm v_x$, ($+$ red, $-$
 green)}
\end{figure}
%%%%%%%%%%%%%%%%%%%%%%%%%%%%%%%%%%%%%%%%%%%%%%%%%%%%%%%
%%%%%%%%%%%%%%%%%%%%%%%%%%%%%%%%%%%%%%%%%%%%%%%%%%%%%%%

\

{\it Spin Orbit and  Effective Magnetic Field -} In
Fig.~\ref{fig4} we report the spin polarization of the transverse
current, when considering a vanishing external magnetic field and
a $\beta$-coupling SO term. Numerical calculations are performed
using the  procedure discussed  above. Thus,  we show the spin
polarization $\langle S_z\rangle$ of the current flowing along the
$x$ direction. $\langle S_z\rangle$ corresponds to
$$
\langle S_z\rangle=\frac{\hbar}{2}\frac{T^{\uparrow
\uparrow}_{21}-T^{\downarrow \downarrow}_{21}}{T^{\uparrow
\uparrow}_{21}+T^{\downarrow \downarrow}_{21}}\equiv \frac{\hbar}{2}
P_z,
$$
in this special case where
$T^{\uparrow\downarrow}_{ij}=T^{\downarrow\uparrow}_{ij}=0$, because
of the commutation between $\hat{S}_z$ and $\hat{H}^\beta_{SO}$ and
$T^{\uparrow\uparrow}_{12}=T^{\downarrow\downarrow}_{14}$,
$T^{\uparrow\uparrow}_{14}=T^{\downarrow\downarrow}_{12}$ since the
effective magnetic field depends on the spin orientation of
electrons injected in 1.

Starting from the number of trajectories we are able to estimate
the statistical fluctuation on the calculated $\langle
S_z\rangle$,
$$\sigma_{S_z}\lesssim \frac{\hbar}{2} \frac{\sigma_B}{N_t}\sim
\frac{\hbar}{2} 0.018, $$ where we take $\sigma_B\propto
\sqrt{N_t}$ according to the binomial distribution.

\

Notice that, for $\varepsilon_y$ between $0.1$ and $0.3 \hbar
\omega_d$,  there is an inversion of the spin polarization
($\langle S_z\rangle<0$) for each panel of Fig. 3. It is well
known that a strong geometry dependence of the transport
properties was shown  in the presence of a transverse magnetic
field by giving negative Hall current, as we discussed above
concerning the "quenching of the Hall effect". In fact, the
resistances measured in narrow-channel geometries are mainly
determined by the scattering processes at the junctions with the
side probes which depend strongly on the junction shape
\cite{ref289}. This dependence of the low-field Hall current was
demonstrated \cite{ref358} and measured \cite{fordbvh}. In a
recent paper\cite{noiq} was discussed how the effective field
generated by the $\beta$-SOC characterizes a regime of transport
that can be assumed as the {\it quenching regime of the SHE}.
Hence, it follows that the inversion of the spin polarization,
shown in Fig.~\ref{fig4}, can be explained on the same ground.
Moreover, this behaviour has a clear signature around
$\varepsilon_y\sim 0.1-0.3 \hbar \omega_d$ while for other values
of the Fermi energy the calculated quenching is comparable with
the statistical fluctuation due to the numerical approach.

The geometry dependence of $\langle S_z\rangle$ can be clearly
inferred by comparing the three panels of Fig. 3, where we show
the effects of the width of the crossing region ($R$). We can
conclude that a significant spin polarization of the transverse
current can be obtained at some fixed values of the Fermi energy,
and that a more efficient process is given by the junction with
$R\sim 4 l_\omega$, while $\langle S_z\rangle$ can be attenuated
in larger or smaller junctions.

By comparing Fig. 2 and Fig. 4, where some trajectories are shown,
we can understand the microscopic mechanism which produces the
transverse spin current, by focusing on the symmetry breaking
between the spin-up and spin-down electrons.

%%%%%%%%%%%%%%%%%%%%%%%%%%%%%%%%%%%%%%%%%%%%%%%%%%%%%%%
%%%%%%%%%%%%%%%%%%%%%%%%%%%%%%%%%%%%%%%%%%%%%%%%%%%%%%%
\begin{figure}
\includegraphics*[width=.95\linewidth]{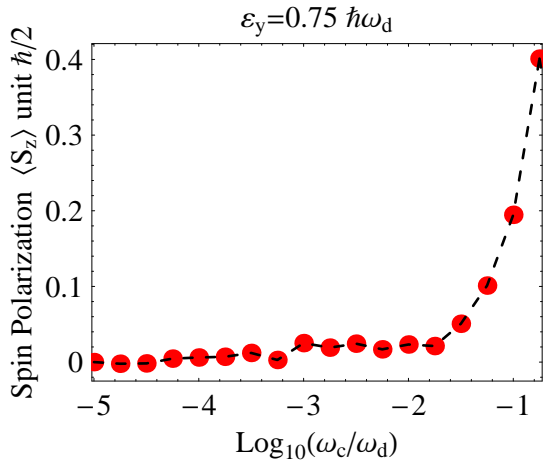}
 \caption {\label{fig2} {\bf Spin polarization at lead 2 $\langle S_z \rangle$ as a function of
the $\beta$ coupling strength  (given by
$\log_{10}(\omega_c/\omega_d)$), for a fixed value of the  Fermi
energy $\varepsilon_F\sim 0.75 \hbar \omega_d$, in the case of
vanishing external magnetic field and $\beta$ SO coupling. } }
\end{figure}
%%%%%%%%%%%%%%%%%%%%%%%%%%%%%%%%%%%%%%%%%%%%%%%%%%%%%%%
%%%%%%%%%%%%%%%%%%%%%%%%%%%%%%%%%%%%%%%%%%%%%%%%%%%%%%%
In order to evaluate the order of magnitude corresponding to the
spin polarization, we can calculate the dependence of $\langle
S_z\rangle$ on the strength of the $\beta$-SOC. In Fig. 5 we show
the value of the spin polarization, as it can be measured at  lead
2, for different strengths of $\beta$ coupling ranging over 5
orders of magnitude.

\

 We
distinguish two cases according to the possibilities that in lead
$1$ is injected, respectively: (i) a non polarized current or (ii)
a polarized charge current. In case (i), there is no charge
current between leads $2$ and $4$, but a pure spin current,
$I_{SH}$, that is proportional to
$T^{\uparrow\uparrow}_{12}-T^{\downarrow\downarrow}_{12}$. This
quantity can also be read as the spin polarization $\langle
S_z\rangle$ of the current in the lead 2, when an unpolarized spin
current is injected in the lead 1. Fig.~\ref{fig4} shows that, for
some energies, this spin current quenches and eventually reverses
its sign. The same is observed for a fixed value of the energy,
when changing the parameter $\omega_c$ (different values of
$\omega_c$ could be experimentally obtained  by changing the value
of the effective width of the wires defining the T-junction, or by
changing the value of the coupling parameter). This phenomenon has
been treated in a recent paper\cite{noiq} studying the transport
through micrometric ballistic 4-probe X-junctions, where it was
found that this magneto-transport anomaly is closely related to
the quenched or negative Hall resistance. In case (ii) a charge
current flows between leads 2 and 4, as can be seen considering a
completely polarized injected current (e.g. having all electrons
with spin up). In that case
$T^{\uparrow\uparrow}_{12}-T^{\uparrow\uparrow}_{14}$ will be
proportional to a charge current flowing between leads 2 and 4. It
is easy to see that, even if the current is not completely
polarized, there will be a charge current flowing between leads 2
and 4 that is proportional to the polarization degree of the
injected current,
$$I_{24}\propto G_{24}=\frac{e^2}{h}\epsilon P_z^0,$$
 where  $G_{24}$ is the charge conductance, $\epsilon\equiv T^{\uparrow\uparrow}_{12}-T^{\downarrow\downarrow}_{12}$ and
$P_z^0$ is the spin polarization of the injected current. This
scheme would implement the electric detection of a spin polarized
current.

\

{\it Conclusion -} We described a system based on the SO
$\beta$-coupling capable of spin filtering and electric based spin
polarization measurements. The results we have shown were obtained
using values of $\lambda$ and $W$ well within those given by
presently available 2DEGs and nanolithography techniques. The use
of a series of T nanojunctions could also give some better results
in the spin polarization of the emerging current.

The proposed devices also represent a new test for the effects of
the  $\beta$-SO interactions which are at the basis of the quantum
spin Hall effects recently discussed in several
papers\cite{noish,qse,iii,hatt,noiq}. {In these papers  the
$\alpha$-coupling was always assumed to be negligible, although in general
this term is comparable to (or larger than) the $\beta$-one. However,
it can be shown that spin polarization effects  of the $\beta$
coupling should be some orders of magnitude larger than the one
calculated for the $\alpha$ coupling with a comparable strength\cite{noiq}. %It must also
%be considered that,  using highly symmetric heterostructures,
%it is possible to strongly reduce $E_z$ and $\alpha$\cite{koga}. }
%%%--------------------------------------------------------

%%%------------------------------------------------------------------------
\bibliographystyle{prsty} %Phys. Rev. style

\bibliography{}

\begin{thebibliography}{99}

\bibitem{spintro} D.D. Awschalom, D. Loss and N. Samarth, {\it Semiconductor Spintronics and Quantum Computation } (Springer, Berlin, 2002);
B. E. Kane et al., Nature {\bf 393}, 133 (1998).

\bibitem{[3]} M. I. D'yakonov and V. I. Perel', JETP Lett. {\bf 13}, 467 (1971).

\bibitem{[7]} J. Sinova, D. Culcer, Q. Niu, N. A. Sinitsyn, T. Jungwirth, and A.
H. MacDonald, Phys. Rev. Lett. {\bf 92}, 126603 (2004).

\bibitem{Hir}
J.~E. Hirsch,  \prl {\bf 83}, 1834 (1999).

\bibitem{she2}
E. M. Hankiewicz, L. W. Molenkamp, T. Jungwirth, and J. Sinova,
Phys. Rev. B {\bf 70}, 241301 (2004).
\bibitem{she3}
C. L. Kane and E. J. Mele, Phys. Rev. Lett. 95, 226801 (2005).
\bibitem{she4}
L. Sheng, D. N. Sheng, C. S. Ting, and F. D. M. Haldane, Phys.
Rev. Lett. {\bf 95}, 136602 (2005).

\bibitem{Thankappan}  L.~D.~Landau, E.~M.~Lifshitz, {\it Quantum Mechanics} (Pergamon Press, Oxford, 1991).

\bibitem{BR} Yu.~A.~Bychkov, E.~I.~Rashba, Pis'ma Zh.~Eksp.~Teor.~Fiz. {\bf
39}, 66 (1984) [JETP Lett. {\bf 39}, 78 (1984)]).

\bibitem{Kelly} M.~J.~Kelly {\it Low-dimensional semiconductors: material, physics, technology, devices} (Oxford University Press, Oxford, 1995).


\bibitem{bonew} S. Bellucci and P. Onorato, Phys. Rev. B {\bf 72}, 045345 (2005); S. Bellucci and P. Onorato, Phys. Rev. B {\bf 68}, 245322 (2003).

\bibitem{shapers} G.~Engels, J.~Lange, Th.~Sch\"apers, and H.~L\"uth, Phys. Rev. B  \textbf{55}, R1958 (1997).

\bibitem{Schultz} M.~Schultz, F.~Heinrichs, U.~Merkt, T.~Colin, T.~Skauli, and S.~L{\o}vold, Semicond. Sci. Technol.  \textbf{11}, 1168 (1996).

\bibitem{1519} S.-Q. Shen, Phys. Rev. Lett. {\bf 95}, 187203 (2005).
\bibitem{1519a} B. K. Nikolic, L. P. Zarbo, and S. Welack, Phys. Rev. B {\bf 72},
075335 (2005).
\bibitem{1519b} B. Zhou, L. Ren, and S.-Q. Shen, Phys. Rev. B {\bf 73}, 165303
(2006).
\bibitem{1519c} A. Berard and H. Mohrbach, Phys. Lett. A {\bf 352}, 190 (2006).

\bibitem{Thornton} T.~J.~Thornton, M. Pepper, H. Ahmed, D. Andrews, G.~J.~Davies, \prl {\bf 56}, 1198 (1986).

\bibitem{morozb} A.~V.  Moroz and C.~H.~W. Barnes, Phys. Rev. B \textbf{61}, R2464 (2000).


\bibitem{noish} S. Bellucci and P. Onorato, Phys. Rev. B {\bf 73}, 045329 (2006).

\bibitem{qse}
B. A. Bernevig and S.-C. Zhang, Phys. Rev. Lett. {\bf 96}, 106802
(2006).
\bibitem{iii}
Y. Jiang and L. Hu, cond-mat/0603755.
\bibitem{hatt}
Kiminori Hattori and Hiroaki Okamoto, Phys. Rev. B \textbf{74},
155321 (2006).
\bibitem{noiq} S. Bellucci and P. Onorato, Phys. Rev. B \textbf{74},
245314 (2006).

\bibitem{kk} A.~A.~Kiselev and K.~W.~Kim, Appl. Phys. Lett.  \textbf{78}, 775 (2001).


%\bibitem{koga} T. Koga, J. Nitta, T. Akazaki, and H. Takayanagi Phys. Rev. Lett. \textbf{89}, 046801 (2002).

\bibitem{3840} S.~E.~Laux, D.~J.~Frank, F.~Stern, Surf.~Sci. {\bf 196}, 101
(1988); H.~Drexler {\it et al.}, \prb {\bf 49}, 14074 (1994); B.
Kardyna\l {\it et al.}, \prb {\bf 55}, R1966 (1997).

%\bibitem{3t} G. Dresselhaus, Phys. Rev. {\bf 100}, 580 (1955).

\bibitem{carillo} F.~Carillo, G.~Biasiol, D.~Frustaglia, F.~Giazzotto, L.~Sorba, F.~Beltram, Phys. E {\bf32}, 53 (2006).



\bibitem{HgTe} X. C. Zhang, A. Pfeuffer-Jeschke, K. Ortner, V. Hock, H. Buhmann, C. R. Becker, and G. Landwehr, Phys. Rev. B {\bf63}, 245305 (2001).

\bibitem{kunze} M. Knop, M. Richter, R. Ma\ss{}mann, U. Wieser, U. Kunze, D. Reuter, C. Riedesel and A. D. Wieck Semicond. Sci. Technol. {\bf20},
814 (2005).
\bibitem{BL} M. B\"uttiker \prl {\bf 57}, 1761 (1986).


\bibitem{gei92} T. Geisel,  R. Ketzmerick and O. Schedletzky, \prl {\bf 69}, 1680 (1992).


\bibitem{fordbvh}
C. J. B. Ford, S. Washburn, M. B\"uttiker, C. M. Knoedler, J. M.
Hong, Phys. Rev. Lett. \textbf{62}, 2724 (1989).




\bibitem{ref289} G. Timp, H. U. Baranger, P. deVegvar, J. E. Cunningham, R. E. Howard, R. Behringer, P. M. Mankiewich, Phys. Rev. Lett. \textbf{60}, 2081 (1988).


\bibitem{ref358} H. U. Baranger, A. D. Stone, Phys. Rev. Lett. \textbf{63}, 414
(1989).



 %%%%%%%%%%%%%%%%%%%%%%%%%%%%%%%%%%%%%%

%%%%%%%%%%%%%%%%%%%%%%%%%%%%%%%%%%%%%%%%%%%%%%%%%%%%%%%%%%%%%%%%%%
\end{thebibliography}

\end{document}